\documentclass[12pt]{iopart}
\usepackage{graphicx}

\begin{document}

\title{Spectral evolution in an insulator exhibiting linear
specific heat}

\author{R. Bindu, Ganesh Adhikary, Sudhir K. Pandey and Kalobaran Maiti}

\address{Department of Condensed Matter Physics and
Materials Science, Tata Institute of Fundamental Research, Homi
Bhabha Road, Colaba, Mumbai - 400 005, INDIA.\\}
\ead{kbmaiti@tifr.res.in}

\begin{abstract}
We investigate the spectral evolution of an antiferromagnetic
insulator, La$_{0.2}$Sr$_{0.8}$MnO$_3$ exhibiting linear specific
heat using state-of-the-art high resolution photoemission
spectroscopy. Experimental spectral functions exhibit Fermi liquid
like energy dependence at all the temperatures studied. Room
temperature spectrum possess finite density of states at the Fermi
level that vanishes generating a soft gap at about 260 K (the
magnetic transition temperature). High resolution spectra reveal a
hard gap in the magnetically ordered phase (C-type antiferromagnet).
These results indicate signature of an amorphous phase coexisting
with the long range ordered phase in these materials.

\end{abstract}


\maketitle

Specific heat of an electron gas has a linear dependence
with temperature and the specific heat coefficient, $\gamma$ can be
expressed as $\gamma = {\pi^2\over 3} k_B^2 n(\epsilon_F)$, where
$k_B$ is the Boltzmann constant and $n(\epsilon_F)$ is the density
of states (DOS) at the Fermi level, $\epsilon_F$. Here, $\gamma$ is
a linear function of $n(\epsilon_F)$. However, various experiments
reported finite value of $\gamma$ in insulating materials such as
vitreous silica, germania, selenium etc \cite{Zeller}. Subsequently,
it was proposed \cite{Anderson, Phillip} that in insulating
materials the charge carriers can tunnel through the potential
barrier among various local minima. The energy difference between
the local minima will vary continuously in an amorphous system that
leads to a linear term in specific heat. Interestingly, various
studies in manganites reveal large $\gamma$ in insulating {\it
crystalline} compositions \cite{salamon, chaddah}. In order to
explain such anomalous observations, it was suggested that the DOS
at $\epsilon_F$ is finite but localized that gives rise to finite
$\gamma$ and insulating transport. The other suggestion is the
possibility of spin glass phase \cite{Ghivelder}.

\begin{figure}
\vspace{-2ex}
\begin{center}
\includegraphics [scale=0.5]{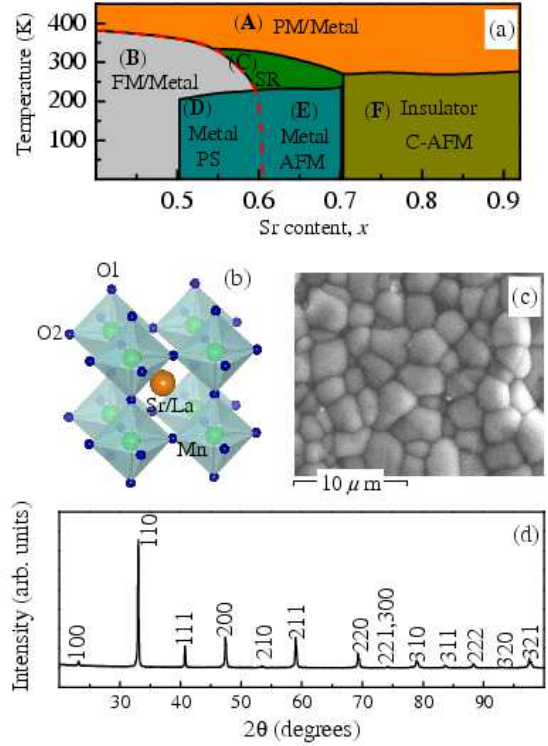}
\end{center}
\vspace{-8ex} \caption{(color online) (a) Phase diagram as a
function of $x$ in La$_{1-x}$Sr$_x$MnO$_3$. The paramagnetic (PM)
metallic phase (A) at high temperatures undergoes to ferromagnetic
(FM) metallic (B) or via short range (SR) ordered phase (C) to
phase-separated (PS) metallic (D) or antiferromagnetic (AFM)
metallic (E) phase, or antiferromagnetic insulating phase (F). (b)
Crystal structure of La$_{0.2}$Sr$_{0.8}$MnO$_3$. (c) Scanning
electron microscopic image of the sample studied exhibiting large
grain size of single phased material. (d) $X$-ray diffraction
pattern of the sample revealing clean high quality phase of the
sample.}
\vspace{-2ex}
\end{figure}

Hole doped manganites \cite{coey,dagotto,salamon,Tokura,CNRao,TVR},
in general, have attracted a great deal of attention during last two
decades followed by the discovery of colossal magnetoresistance
(CMR) leading to potential technological applications. In addition,
plethora of interesting phases are observed due to the interplay
between spin, charge, orbital and lattice degrees of freedom. While
CMR effect observed at low doping levels, the higher doping regime
also displays interesting and complex phase diagram as shown in Fig.
1(a) in the case of La$_{1-x}$Sr$_x$MnO$_3$
\cite{Hemberger,Chmaissem}. Clearly, the system lies in the
proximity of ferromagnetic and antiferromagnetic ground states;
phase coexistence is observed in the composition range (0.5 $< x <$
0.6). Thus, such systems may lead to spin amorphicity that
contributes as finite $\gamma$ in the specific heat data.

However, the observation of finite $\gamma$ in the insulating
materials far away from compositions having phase coexistence is
curious. For example, La$_{0.2}$Sr$_{0.8}$MnO$_3$ undergoes a
transition from paramagnetic metallic phase to C-type
antiferromagnetic insulating phase (moments in $ab$-plane are
antiparallel and inter-plane coupling is ferromagnetic). $\gamma$ in
La$_{0.2}$Sr$_{0.8}$MnO$_3$ is $\sim$5.6 mJ/(mole.K)
\cite{Bindu_APL}, which is very close to the values observed in
metallic compositions \cite{salamon}. The crystal structure is
perovskite derived as shown in Fig. 1(b). Here, we report our
results on La$_{0.2}$Sr$_{0.8}$MnO$_3$ employing high resolution
photoemission spectroscopy. We find that DOS at $\epsilon_F$ is
finite at room temperature although the $e_g$ electrons possess
strong local character. Antiferromagnetic transition leads to an
energy gap at $\epsilon_F$. The spectral function exhibits Fermi
liquid like energy dependence at all the temperatures studied (even
in the gapped phase).

The samples were prepared by solid-state reaction route as reported
elsewhere \cite{BinduEPJB}. Scanning electron microscopic (SEM)
picture [see Fig.1(c)] reveals large grain size ($\sim$~3~${\mu}m$)
that could be achieved by long sintering at the final preparation
temperature. The energy dispersive analysis of $x$-rays on different
grains and different location on same grain indicate absence of
impurity phase and homogeneity of the composition. The room
temperature powder $x$-ray diffraction (XRD) experiment was carried
out using PHILIPS X'Pert diffractometer with Cu K$\alpha$ radiation.
All the reflections were indexed with cubic structure (space group
$Pm\bar{3}m$) with lattice parameter of 3.826 \AA, (see Fig 1(d)).
No trace of impurity is found in the XRD pattern. The dc
magnetization measurements (4 - 330K), carried out at 5 Tesla field
in a superconducting quantum interference device (Quantum design)
exhibits a distinct hump at about 265 K [see Fig. 2(a)] indicating
transition to antiferromagnetic phase.

\begin{figure}
 \vspace{-6ex}
 \begin{center}
\includegraphics [scale=0.6]{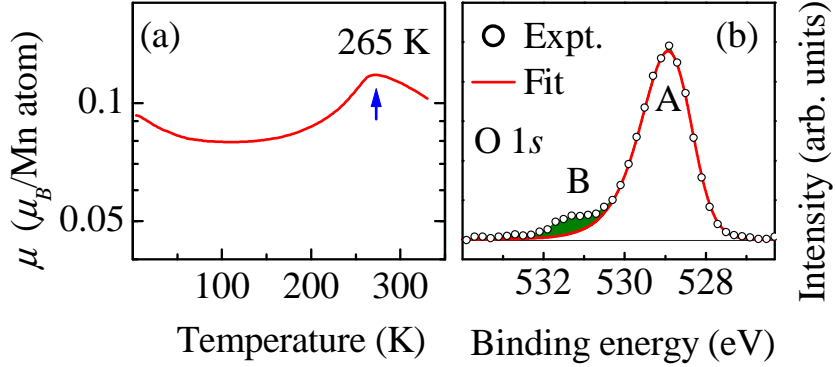}
\end{center}
\vspace{-60ex}
 \caption{(color online) (a) Magnetic susceptibility as a
function of temperature exhibiting antiferromagnetic transition at
265 K. (b) O 1$s$ spectrum exhibiting high purity of the sample.}
 \vspace{-2ex}
\end{figure}

The photoemission measurements at different temperatures were
carried out using monochromatic Al K$\alpha$ ($h\nu$ = 1486.6 eV),
He {\scriptsize I} ($h\nu$ = 20.2 eV) and He {\scriptsize II}
($h\nu$ = 40.8 eV) sources, and Gammadata Scienta analyzer SES2002.
The energy resolution for $x$-ray photoemission (XP),
He~{\scriptsize I} and He~{\scriptsize II} measurements are 0.3 eV,
1.4 meV and 4.2 meV, respectively. The base pressure during the
measurements was 4 $\times$ 10$^{-11}$ Torr. The sample surface was
cleaned by scraping in-situ with a diamond file. No intensity was
observed for C 1$s$ signal. A typical O 1$s$ spectrum collected at
room temperature is shown in Fig 2(b) exhibiting a sharp feature, A
at 529 eV binding energy and a weak feature, B at about 531.3 eV
binding energy. The feature B can be attributed to the surface
oxygens and/or adsorbed impurities. The complete dominance of the
feature A ensures high purity of the samples.

\begin{figure}
 \vspace{-2ex}
 \begin{center}
\includegraphics [scale=0.42]{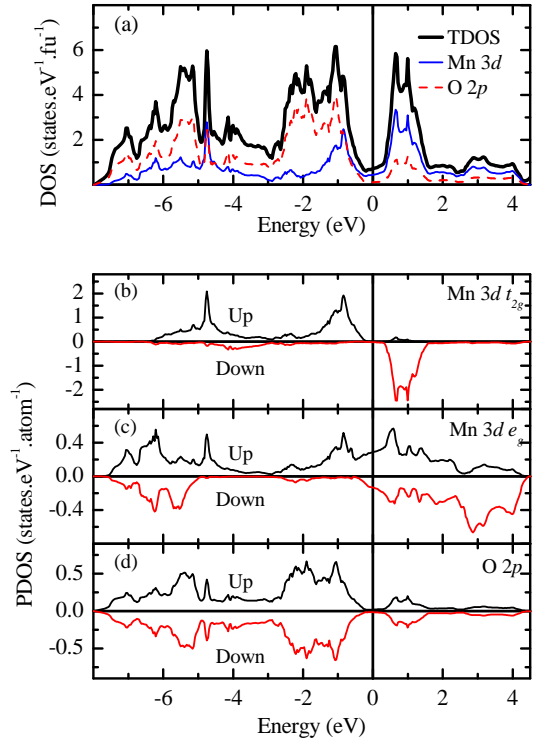}
\end{center}
\vspace{-8ex}
 \caption{(color online) (a) Calculated spin integrated density of states
for the C-type antiferromagnetic phase. Thick line, thin line and
dashed line represent the total DOS, Mn 3$d$ PDOS and O 2$p$ PDOS,
respectively. Spin-resolved density of states are shown for Mn 3$d$
PDOS with (b) $t_{2g}$ symmetry and (c) $e_g$ symmetry, and (d) O
2$p$ PDOS.}
 \vspace{-2ex}
\end{figure}

The band structure calculations were carried out by using a
linearized muffin-tin orbital method within the atomic sphere
approximation (LMTART 6.61) \cite{Savrasov}. The muffin-tin radii
used are 3.523, 3.523, 2.025, 1.591 a.u. for La, Sr, Mn and O
respectively. The charge density and effective potential were
expanded in spherical harmonics up to $l$= 6 inside the sphere. The
exchange correlation functional of the density functional theory was
taken after Vosko $\emph{et al}$ \cite{Vosko}. The convergence in
the total energy was set to 10$^{-5}$ Ryd/cell. (8,8,8) divisions of
the Brillouin zone along three directions for the tetrahedron
integration were used to calculate DOS. The calculations were
performed for La$_{0.25}$Sr$_{0.75}$MnO$_{3}$ in the C-type
antiferromagnetic phase.

The calculated DOS corresponding to the valence band are shown in
Fig. 3. Mn 3$d$ states split into $t_{2g}$ and $e_g$ bands due to
the crystal field of MnO$_6$ octahedra. The bonding and antibonding
$t_{2g}$ up spin bands are centered around -5 eV and -1 eV,
respectively; the energy separation is about 4 eV. Almost equal
weight of the $t_{2g}$ partial DOS in bonding and antibonding bands
indicates its strong mixing with the O 2$p$ electronic states. The
down spin partial DOS appear above the Fermi level as expected for
Mn$^{3+}$/Mn$^{4+}$ high spin state. The bonding $e_g$ bands appear
between -4 to -8 eV. The contribution at the Fermi level,
$\epsilon_F$ arises primarily due to the up spin $e_g$ electronic
states. O 2$p$ partial DOS has large contributions in the bonding
and antibonding energy regions. The non-bonding O 2$p$ contributions
appear in the energy range of -1 to -3 eV.

The experimental valence band spectra are shown in Fig 4. There are
three intense discernible features A, B and C in both XP and He
{\scriptsize II} spectra. First, we discuss the observations in the
room temperature spectra. The feature A in the XP spectrum appears
at slightly higher binding energy compared to that in the He
{\scriptsize II} spectrum, while the features B and C appear to have
similar binding energies. The feature C intensity is reduced
slightly in the He {\scriptsize II} spectrum compared to that in the
XP spectrum. The contributions from Mn 3$d$ states are dominant in
the XP spectrum and that from O 2$p$ states are dominant in the He
{\scriptsize II} spectrum due to photoemission cross section. Thus,
these observations suggest strongly mixed character of the features
along with a small enhancement in Mn 3$d$ contribution in feature C.
Comparing these results with the calculated DOS, it is evident that
the feature A represents the photoemission signal from bonding bands
(both $t_{2g}$ and $e_g$), the feature B is the non-bonding O 2$p$
contributions and the feature C is the antibonding $t_{2g}$ bands.
The peaking of feature A at slightly higher binding energy in the XP
spectrum presumably due to the enhanced intensity of the bonding
$e_g$ bands contributions at XP energies. In addition, a weak
feature D can also be observed (see inset) in the vicinity of the
Fermi level, which is dominant in the XP spectra and represent the
electronic states having $e_g$ symmetry as evident in Fig. 3(c). All
these spectra exhibit negligible intensity at the Fermi level
suggesting proximity to an insulating phase. This is consistent with
the results from resistivity measurements.

\begin{figure}
 \vspace{-6ex}
 \begin{center}
\includegraphics [scale=0.55]{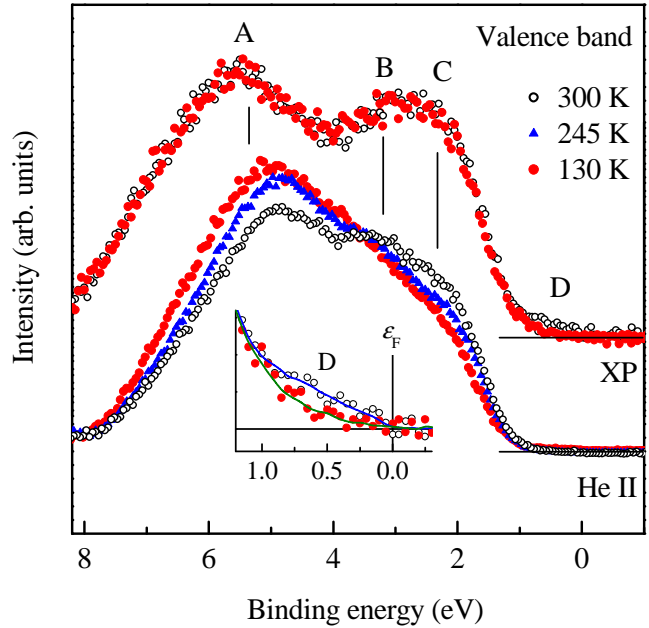}
\end{center}
\vspace{-30ex}
 \caption{(color online) XP and He {\scriptsize II} valence band spectra
at different temperatures. The inset shows the rescaled XP spectra
in the vicinity of $\epsilon_F$.}
 \vspace{-2ex}
\end{figure}

The energy and intensity of the features in XP spectrum remain
almost identical down to 130 K (much below the magnetic transition
temperature of 265 K). On the other hand, the He {\scriptsize II}
spectra (normalized by the intensity of non-bonding O 2$p$ signal)
exhibit significant spectral weight transfer. The feature A becomes
stronger at lower temperatures with subsequent reduction in
intensity of feature C. Since the O 2$p$ character is dominant in
the He {\scriptsize II} spectra, these spectral evolution suggests
the shift of O 2$p$ eigen energies towards higher binding energies.
The antiferromagnetic coupling among Mn $t_{2g\uparrow}$ moments in
the $ab$-plane is mediated by O 2$p$ electronic states
(superexchange interaction). It appears that the onset of
antiferromagnetic ordering leads to a higher degree of localization
of the O 2$p$ electrons.

\begin{figure}
 \vspace{-4ex}
 \begin{center}
\includegraphics [scale=0.6]{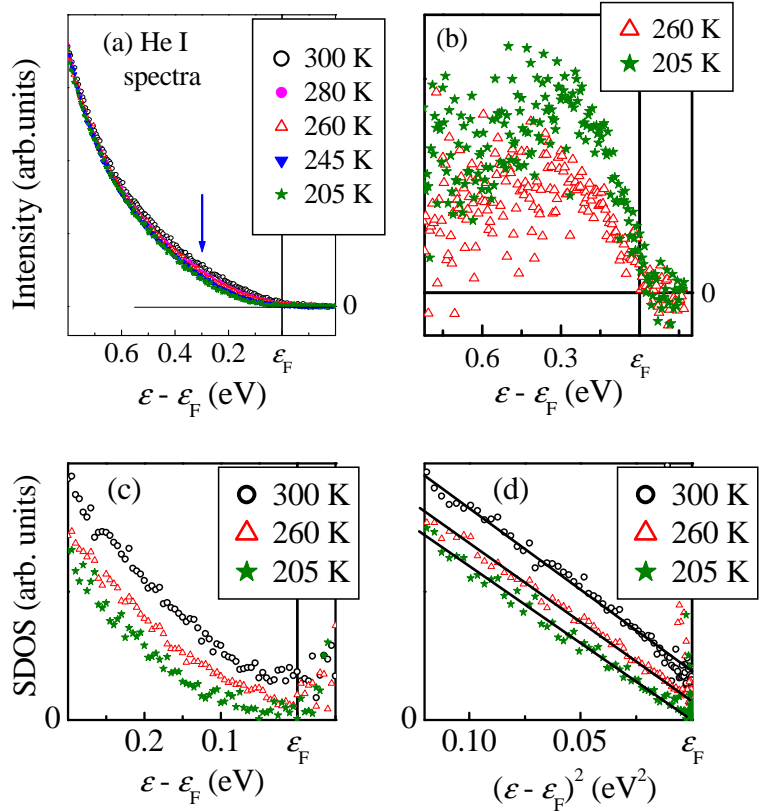}
\end{center}
\vspace{-24ex}
 \caption{(color online) (a) He {\scriptsize I} spectra at different
temperatures. (b) Difference spectra at 260 K and 205 K from the
room temperature spectrum. (c) Spectral density of states (SDOS)
obtained by dividing the spectra by the Fermi-Dirac function. (d)
SDOS plotted as a function of $(\epsilon - \epsilon_F)^2$ revealing
Fermi liquid like energy dependence.}
 \vspace{-2ex}
\end{figure}

The inset of Fig. 4 reveals significant reduction of intensity of
the $e_g$ band (feature D) indicating an opening of a hard gap in
the antiferromagnetic phase. In order to investigate this with
better clarity, we probed this energy region with very high energy
resolution of 1.4 meV. The spectra collected at different
temperatures are shown in Fig. 5(a). Evidently the intensity at
$\epsilon_F$ in the room temperature spectrum is weak and a change
in intensity is observed near the arrow. To visualize the spectral
changes, we have subtracted all the spectra from the room
temperature spectrum. The subtracted intensities shown in Fig. 5(b)
exhibit a peak at about 0.3 eV which enhances gradually with the
increase in temperature difference. Qualitatively, it is expected
that 0.2 electrons in the $e_g$ bands would pin the Fermi level at
the lower half of the conduction band as seen in Fig. 3(c). However,
the experimental results suggests that these electrons have
significant local character that leads to the peak at 0.3 eV and
weak intensity at $\epsilon_F$. Interestingly, the spectral
intensity of this local feature shifts towards higher binding
energies at lower temperatures as also evidenced in the XP spectra
shown in the inset of Fig. 4.

The spectral density of states (SDOS) are extracted by dividing the
experimental spectra by the corresponding Fermi-Dirac distribution
function. The SDOS at room temperature [see Fig. 5(c)] exhibits a
dip at $\epsilon_F$ (pseudogap) which leads to a soft gap at 260 K
and a hard gap in the antiferromagnetic phase. It is to note here
that the Ne\'{e}l temperature, $T_N$ marks the onset of the first
order phase transition in this compound, which involves nucleation
and growth process. Hence, pseudogap can be attributed to the
signature of gapped low temperature phase nucleated above $T_N$. The
band gap below $T_N$ can be attributed to the relocation of the
Brillouin zone boundary due to antiferromagnetic ordering in the
$ab$-plane. In addition, the hard gap and vanishing of 0.3 eV
features indicate that the $e_g$ electrons become more localized and
shifts to higher binding energies in the magnetically ordered phase.
A plot of the SDOS as a function of $(\epsilon - \epsilon_F)^2$
exhibit a linear dependence in both paramagnetic and
antiferromagnetic phases indicating Fermi-liquid like behavior of
the localized electrons/quasiparticles.

All the above results establish that the low temperature phase is
insulating due to finite energy gap at $\epsilon_F$ although
$\gamma$ is finite and large (similar to that in the metallic phase
of these compounds). Since, other low energy excitations involving
electron-phonon, electron-magnon etc. do not contribute in the
linear term of specific heat, it is clear that some kind of glassy
phase/amorphicity is present in this systems. The issue of phase
separation and its implication in the colossal magnetoresistance in
these systems are widely discussed \cite{Moreo,TVRa}. It was also
suggested that the magnetic transition is accompanied by the
formation of a pseudogap phase \cite{Mannela} similar to that
observed in high temperature superconductors \cite{kanigel}. Thus,
it is tempting to correlate the behavior of this compound with the
ones showing precursor effects, phase separations etc. These results
emphasizes the need to consider a phase that has sufficient
amorphicity and/or softness \cite{littlewood} coexisting with
antiferromagnetic insulating phase that may contribute as a linear
term in the specific heat. This is not unrealistic as $e_g$
electrons are localized and hence, 20\% La concentration at the Sr
sites in LaSrMnO$_3$ will naturally induce disorder.

In summary, we have investigated the evolution of the electronic
structure of La$_{0.2}$Sr$_{0.8}$MnO$_3$ with temperatures using
high resolution photoemission spectroscopy to probe the origin of
linear term in the specific heat in its insulating phase. We observe
interesting change in the oxygen 2$p$ bands contributing in the
valence band spectra. Spectral intensity is finite at the Fermi
level in the paramagnetic phase and the electronic states seems to
have dominant local character (peak of intensity appears around 0.3
eV). The spectral functions close to the Fermi level exhibit opening
of a hard gap in the antiferromagnetic insulating phase via the
formation of a soft Coulomb gap at the antiferromagnetic transition
temperature. The energy dependence of the spectral function is Fermi
liquid like in all the phases. These results indicate possibility of
an amorphous phase involving localized quasiparticles within the
long range ordered phase.

\end{document}